\documentclass{llncs}

\usepackage[utf8]{inputenc}
\usepackage{booktabs}
\usepackage{tcolorbox}
\usepackage{amsmath}
\usepackage{nccmath}
\usepackage{tabularx}
 \usepackage{graphicx}
\usepackage{amsmath}
\usepackage{multicol,tabularx}
\usepackage{color, colortbl}
\definecolor{lightgray}{gray}{0.9}
\definecolor{Gray}{gray}{0.9}
\usepackage{array}
\usepackage{makecell}
\usepackage{multirow}
\usepackage{csvsimple}

\usepackage{xcolor}
\usepackage[framemethod=tikz]{mdframed}

\definecolor{cccolor}{rgb}{.67,.7,.67}

\usepackage{lineno}
\modulolinenumbers[5]
\usepackage{listings}
\usepackage{lineno}
\modulolinenumbers[5]
\usepackage[T1]{fontenc} 
\usepackage[lighttt]{lmodern}  
\usepackage{newfloat} 
\usepackage{listings}
\DeclareFloatingEnvironment[fileext=lst,placement={htbp},name=Listing,within=none]{lstfloat} 

\newcommand{\head}[1]{\par\noindent\textbf{#1:}}

\title{Invariant-based Program Repair\thanks{Accepted for publication in the 27th International Conference on Fundamental Approaches to Software Engineering (FASE 2024).}}

\author{ Omar I. Al-Bataineh}
\institute{Simula Research Laboratory, Oslo, Norway\\ \email{omar@simula.no}}
\date{}

\begin{document}

\maketitle

\begin{abstract}

This paper describes a formal general-purpose automated program repair (APR) framework based on the concept of program invariants.
In the presented repair framework,  the execution traces of a defected program are dynamically analyzed to infer 
specifications $\varphi_{correct}$ and $\varphi_{violated}$, where  $\varphi_{correct}$ represents the set of likely
invariants (good patterns)  required  for a run to be successful and $\varphi_{violated}$ 
represents the set of likely suspicious invariants (bad patterns) that result in the bug in the defected program.
These specifications are then refined using rigorous program analysis techniques, 
which are also used to drive the repair process towards feasible patches and assess the correctness of generated patches. 
 We demonstrate the usefulness of leveraging invariants in APR
by developing an invariant-based repair system for performance bugs.
The initial analysis shows the effectiveness of invariant-based APR in handling performance bugs by
producing patches that ensure program's efficiency improvement without adversely impacting its functionality.

\end{abstract}

\begin{keywords}
    Automated program repair $\cdot$
    Invariant learning and refinement $\cdot$
    Patch overfitting $\cdot$
    Program verifier $\cdot$
    CPAChecker $\cdot$
    Performance bugs

\end{keywords}

\section{Introduction}

 Automated program repair (APR) has recently gained great attention  because it helps to significantly decrease manual debugging effort by automatically generating patches for defected programs. 
Modern program repair tools have been shown to be effective at fixing bugs in many real-world programs. 
The poor quality of automatically generated patches \cite{QiLAR15}, however, continues to be a major obstacle to the adoption
of automated program repair by software practitioners.

\head{Problem}
The primary reason for the low quality of automatically
generated patches by current APR tools is the lack of specifications of the intended behavior. 
Most program repair systems rely on tests as the correctness
criteria, because a formal specification is not explicitly provided by software developers.
Therefore, current APR approaches produce plausible patches 
which must be (manually) inspected before being deployed. 
Therefore, there is no guarantee that the generated patches are generally correct and do not introduce new bugs.

\head{Solution} 
 Program verification technology enables developers to prove the correctness of the program before deploying it.
One of the key activities underlying this technology involves inferring a program invariant—a logical
formula that serves as an abstract specification of a program.
Developers can significantly benefit from  program invariants
to identify program properties that must be preserved when
modifying code. Unfortunately, these invariants are typically
absent from code,  leading to the dominance of less rigorous APR approaches (e.g., dynamic APR)
and the well-known patch overfitting challenge ~\cite{QiLAR15}.

We argue that by using test cases and reachability-based analysis techniques, an accurate set of invariants may be obtained and utilized to produce high-quality patches. 
In other words, program verification tools such as CPAChecker \cite{Dirk11} and PathFinder \cite{VisserHBPL03} can be used to refine the dynamically generated invariant candidates.
This can be done by first using the test cases to analyze the execution traces of the program to infer a set of invariant candidates. 
These candidates are then refined using a program verifier to obtain more accurate invariants.
The  goal is to infer two specifications: (i) $\varphi_{correct}$, which represents the set of  \textit{good patterns} 
required for a run to succeed, and (ii) $\varphi_{violated}$, which represents the set of \textit{bad patterns} that lead to the target bug. 
Invariant-based APR offers two key benefits.
First, it directs APR  towards potentially feasible patches.
Second, it enables the formal validation of plausible patches using program verifiers.

\head{Viability of invariant-based APR}
Program invariants have shown effectiveness in many applications, such as program understanding, fault localization, and formal verification. 
Invariants are effective  because functional correctness relates to the final
result of a program rather than any specific implementation. 
They can therefore assist in abstracting many concrete execution steps
and thus greatly reduce the effort needed to reason about the patch's correctness.

 In fact, developers who aim to repair a defected \textit{undocumented program} (a program written without thought for formal specifications)
 can find invariant-based APR helpful in their repair tasks. 
 The availability of mature automated invariant detection tools like Daikon \cite{ErnstPGMPTX07}
 and   practical software verification tools like CPAChecker and PathFinder
 makes the invariant-based program repair technique viable.
At first glance, refining invariants using program verification tools seems too expensive. 
However, due to tremendous advances in software verification \cite{Beyer19}, in practice, invariant-based verification can be made pretty efficient. 
In particular, the software analysis framework CPAChecker, which supports many different reachability analyses, 
has been effectively used to validate a wide variety of reachability queries against C programs with up to 50K lines of code. 
This makes reachability analysis a promising technique that can be used to significantly reduce the patch overfitting problem and produce high-quality patches.

\section{Invariant-based Program Repair Framework}

In this section we reformulate the APR problem using the concept of program invariants.
We then describe how one can analyze the execution traces of fault-free runs to infer likely specifications of the program's intended behaviour  
and execution traces of faulty runs to infer likely suspicious invariants that lead to the faulty behaviour. 
Before proceeding further, let us introduce some definitions.

\begin{definition} (\textbf{fault-free vs. faulty runs})\label{SuccVsUnsuccRuns}.
Let $P$ be a buggy program, $\mathcal{R}$ be the set of runs of $P$,
and $\varphi_{beh}$ be a property of  program $P$'s intended behavior. 
We say that a run $r \in \mathcal{R}$ is a successful run (i.e., fault-free run) if
$P (r) \models \varphi_{beh}$.
On the other hand,  we say that a run $r' \in \mathcal{R}$ is a faulty run if $P (r) \not\models \varphi_{beh}$.
\end{definition}

From Definition \ref{SuccVsUnsuccRuns} we note that by analyzing
information extracted from fault-free runs, one might be able 
to infer a specification of the program's intended behavior.
Similarly, by analyzing the execution information of faulty runs, one might be able 
to deduce the violating invariants that cause the bug. 
This is because fault-free runs represent  runs
in which program invariants are maintained, while faulty runs represent runs in which some program invariants are violated.

\begin{definition} (\textbf{Invariant-based APR problem})\label{InvDef}.
Let $P$ be a program containing bug $b$ and $T = (T_P \cup T_F)$ be a test suite, where $T_P $ represents the set of passing tests and $T_F$ represents the set of failing tests.
Let $D$ be a dynamic  invariant inference tool like Daikon, and $V$ be a program verification tool like CPAChecker.
The invariant-based APR process consists of the following steps:

\begin{enumerate}
    \item $[$Invariant extraction$]$. Generate an initial set of invariants  $\mathcal{I}$  for P using $D$.

    \item $[$Invariant refinement$]$. Refine the set  $\mathcal{I}$ using $V$
    to produce specifications $\varphi_{correct}$ and $\varphi_{violated}$. 
    This can be done by asserting invariants at a program's location of interest and using any generated counter-example to refine them.  
    \item $[$Fault localization$]$.  Compute a list of suspicious statements whose mutation may lead to a valid patch by analyzing
       specifications $\varphi_{correct}$ and $\varphi_{violated}$.

    \item $[$Patch generation$]$. Construct code that corrects the invariants that are violated while maintaining other program invariants.
   This can be performed by employing a patch generation procedure like search- or semantic-based. 

    \item $[$Patch validation$]$.  Validate the correctness of the generated patches using $V$. 
\end{enumerate}
    
\end{definition}

Depending on the type of the bug being fixed and the structure of the analyzed program, different program locations may be of relevance for properties $\varphi_{correct}$ and $\varphi_{violated}$.
 Examples include pre- and post-conditions for different functions, or loop invariants for some program loops.
Note that the first two steps of the invariant-based APR process described at Definition \ref{InvDef} are necessary for increasing confidence in the precision of patches that are generated.
The actual repair steps of the process, steps 3-5, can be formally stated as follows:
\begin{equation}
    \begin{medsize}
pt = FV(PGV(FL(\varphi_{correct}, \varphi_{violated}, P), T), \varphi_{correct}, \varphi_{violated})
 \end{medsize}
\end{equation}
where $FL$ is an invariant-based fault localization process, $PGV$ is patch generation and validation process using test suite,
and $FV$ is a formal patch validation process using the verification tool $V$. 
If no plausible patch is found or a plausible patch is found but incorrect, the repair process returns $\mathsf{fail}$. 
However, if the plausible patch passes the verification step carried out by the tool $V$, the process returns a patch.
We now turn to discuss how one can generate specifications $\varphi_{correct}$ and $\varphi_{violated}$   
by analyzing the execution information obtained by running program $P$ using  passing and failing tests.
The analysis  of fault-free and faulty runs leads to the identification of the following formal patterns.
\begin{enumerate}
    \item $\varphi_{correct} = \mathcal{I}_{good} = V (D(P, T_P))$, invariants deduced using only successful runs. This set of invariants represents the likely intended behavior of $P$.

    \item $\varphi_{faulty} =  \mathcal{I}_{mix} = V(D(P, T_F))$, invariants deduced using the set of faulty runs. 
        Note that  the set $ \mathcal{I}_{mix}$ may contain both good and bad patterns depending on how the target bug affects different functionalities of $P$.
    
    \item  $\varphi_{violated} = ( \mathcal{I}_{mix} \setminus  \mathcal{I}_{good})$, the set of violated invariants related to the bug.
    
\end{enumerate}

It is important to categorize and distinguish inferred patterns (invariants) into good and bad patterns, especially when dealing with programs that have several functional requirements.
This helps to identify the set of desired invariants to be maintained and violated invariants to be repaired when modifying code.
It also helps to identify the set of invariants that are relevant to the analyzed bug.
The soundness of inferred $\varphi_{correct}$ and $\varphi_{violated}$ 
depends heavily on the soundness of the employed invariant inference tool as well as the invariant refinement process.
Increasing the amount of program behavior exercised using reachability analysis
increases the likelihood that $\varphi_{correct}$ and $\varphi_{violated}$ are true.

\begin{definition} (\textbf{Patch validation in invariant-based APR}).
Let $P$ be a program containing bug $b$ and $T$ be a test suite containing at least one failing test and one passing test.
Let also $pt$ be a plausible patch that makes $P$ passes all test cases in $T$.
The validity of patch $pt$ can be formally checked as follows
\begin{equation} \label{CorrSpec}
    \begin{medsize}
validity (pt) = V (pt, \varphi_{correct}) \land \neg V (pt, \varphi_{violated})
\end{medsize}
\end{equation}
where $ V (pt, \varphi_{correct}) \in \{true, false\}$  and that the tool's response depends on whether the specification is fulfilled or violated in the program being examined. 
\end{definition}

To boost confidence in the validity of the resulting patch, 
we opt to check patches against both $\varphi_{correct}$ and $\varphi_{violated}$.
However, to lower the cost of calling the  verifier $V$ against each candidate patch,
we aim to implement a three-step patch validation method that uses the test suite first 
and the program verifier afterwards.
Generating plausible patches is done in the first step using test cases.
Second step involves formally checking plausible patches against the set of bad patterns (property $\varphi_{violated}$).
Patches that pass the first two steps are  checked against the set of good patterns (property $\varphi_{correct}$) in the third step. 

 \section{Fixing Performance Bugs Using Invariant-based APR}
 Performance bugs are programming errors that cause significant performance degradation - lead to low system throughput.
Experience has shown that many commercial software that is widely used suffer from performance problems  \cite{SongL17,JinSSSL12,Adrian2013}.  
Therefore, there is a need to develop a rigorous repair framework for performance bugs 
that ensures efficiency gain without compromising functionality.

One unique characteristic of performance bugs comparing to functional bugs is that 
performance bugs do not affect the functionality of the program (i.e., the program is \textit{semantically correct but inefficient}) 
and thus the intended behavior of the program can be automatically deduced using an invariant inference tool. 

This section describes an invariant-based APR system for performance bugs and demonstrates 
how it may be applied to handle performance bugs by producing
patches that ensures efficiency
improvement without sacrificing functionality.

\subsection{Invariant-based Repair Framework for Performance Bugs}
In this section we describe an invariant-based repair framework for handling performance bugs.
The framework consists mainly of the following components:

\begin{enumerate}
\item  a set of passing tests (tests that lead to fast runs), 
\item a set of failing tests (tests that lead to slow runs), 
 \item  runtime monitor to keep track of the program's execution time and differentiate between fast and slow runs, and
 \item an automated invariant inference tool (Daikon or CPAChecker) and automated invariant verification tool (PVS, Z3 solver, or CPAChecker). 
\end{enumerate}

We now turn to discuss how we define the notions of passing and failing tests 
and the process of generating and validating patches for performance bugs. 

 \head{Passing and failing tests for performance bugs} 
Performance bugs  do not  produce debugging information at runtime: they do not produce  crashes, exceptions, or incorrect results.
We therefore use a runtime monitor with a predefined timer to redefine the concepts of passing and failing tests. 
We consider test cases that lead to \textit{fast runs} as passing tests 
while test cases that lead to \textit{slow runs}  as failing tests.
A repair that transforms slow runs into fast runs while preserving the desired behavior of the original program is considered as a valid repair.
    
\head{Patch generation strategy for performance bugs}
Since we deal with a semantically correct but inefficient program, an efficient version of the program can often be created by restructuring the original program's basic components.
Our preliminary analysis demonstrates the effectiveness of genetic repair tools, such as GenProg, in dealing with performance bugs.
This suggests that programs with performance bugs can be fixed by relatively simple changes.
For instance, various performance bugs can be fixed by using mutation operators like move, swap, delete, and insert employed by genetic repair programs. 
Consequently, we aim to combine our repair framework with genetic-based patch generation tools.

\head{Patch validation for performance bugs}
It should be noted that invariant inference tools can also be used to derive predicates related to the non-functional attributes of the program.
This can be achieved by adding extra non-functional variables to the program being repaired. 
Suppose we have a program $P$ with a set of variables $V$ and that $P$ containing a performance bug.  
We need to check whether the generated plausible patch for program $P$ fixes the performance bug without introducing new functional bug. 
To do so, we first generate and validate predicates related to the efficiency attributes of the program, as described below.

\begin{enumerate}
    \item Add a fresh variable \verb+nfv+ whose value has no impact on the behavior of  $P$. 
    The type of performance bug that is being handled determines how \verb+nfv+ is used to model the efficiency of the program.
    However, for the loop programs we consider, \verb+nfv+  acts like a counter that is incremented once for each iteration.
In other words, the number of loop iterations serves as a model for efficiency.
    \item Use the invariant detection tool $D$ to infer the numerical invariants $\mathcal{I} (P,  ~$\verb+nfv+$ )$ and $\mathcal{I} (pt,  ~$\verb+nfv+$ )$
    for the original and plausible patched version, where $\mathcal{I} (P,  ~$\verb+nfv+$ )$ represents the collection of invariants in program $P$ involving variable  \verb+nfv+. 

    \item Compare the numerical predicates in   $\mathcal{I} (P,  ~$\verb+nfv+$ )$ and $\mathcal{I} (pt,  ~$\verb+nfv+$ )$  
          to determine whether the patched version $pt$ is more efficient than original program $P$.

\end{enumerate}

For simplicity reasons, we assume we deal with a program with a single loop.
The number of loops in the analyzed program, however, determines how many more variables are needed. 
The invariant inference tool $D$ is thus used to infer invariants on $(V \cup \{$\verb+nfv+\}$)$.
We then distinguish the following types of predicates:
\begin{itemize}
   \item $\mathcal{I} (P,  V)$:  predicates related to the program's functionality, and

   \item $\mathcal{I} (P,  $~\verb+nfv+$)$:  predicates related to the program's efficiency.
\end{itemize}  
Using the generated  predicates,  one can check the validity of patch $pt$ as follows
\begin{equation} \label{MainSpec}
        \begin{medsize}
validity (pt) = \textsc{SemaEq} ~
(\mathcal{I} (P,  V ), \mathcal{I} (pt, V)) ~ \land \textsc{PredSm} ~ (\mathcal{I} (pt,  ~\verb+nfv+), \mathcal{I} (P,  ~\verb+nfv+) )
\end{medsize}
\end{equation}
where $\textsc{SemaEq}$ is a Boolean operation that checks whether the given sets of invariants are semantically equivalent
and $\textsc{PredSm}$ is a Boolean operation that checks whether the upper bound in the  predicate related to the patched version is smaller than the upper bound in the  one related to the original program.

We now describe two formal procedures to verify the validity of plausible patches (specification (\ref{MainSpec})) 
using the available program verification tools.

\begin{enumerate}
    \item \textit{Daikon-PVS}: In this patch validation procedure, Daikon is used to generate predicates related to the functional and efficiency attributes of programs $P$ and $pt$. 
                      In the event that $\mathcal{I} (P,  V)$ and  $\mathcal{I} (pt,  V)$ (i.e., predicates related to functional attributes) are not identical, it may be necessary to examine both equivalence and implication relations between the predicates in those sets in order to determine whether $P$ and $pt$ are semantically equivalent. By querying the theorem prover PVS, this task can be accomplished. 


      \item \textit{CPAChecker-PVS}: One interesting feature in CPAChecker is that it
      produces correctness witnesses in GraphML format and in those witnesses, one can find the invariants of the analyzed program.
      This feature can be utilized to generate the set of invariants in both the original program
      and corresponding plausible one. In case that the invariants generated for both programs are not identical, 
      it may be necessary to examine both equivalence and implication relations between the predicates in the two sets by invoking the prover PVS.

\end{enumerate}

 
\subsection{Fixing  real-world performance bugs using invariant-based APR}

In this section, we show how invariant-based APR can be used to handle real-world performance bugs.
For space reasons, we only consider one interesting example of performance bugs (see Listing \ref{challengingEx}). 
The bug is based on a real-world flaw that occurred in Apache and has also been analyzed by other researchers~\cite{song2017:performance}.

\begin{lstfloat}[t]
\begin{lstlisting}
int found = -1; 
while (found < 0 ) {
  // Check if string source[] contains target[]
  char first = target[0];
  int max = sourceLen - targetLen;
  for (int i = 0; i <= max; i++) { 
    // Look for first character. 
    if (source[i] != first) {
      while (++i <= max && source[i] != first);
    }
    // Found first character 
    if (i <= max) {
      int j = i + 1; 
      int end = j + targetLen - 1;
      for (int k=1; j<end && source[j]==target[k]; j++, k++);
      if (j == end) {
        /* Found whole string target. */
        found = i;
        break;
      }
    }
  }
  // append another character; try again 
  source[sourceLen++] = getchar();
}
\end{lstlisting}
\caption{A challenging performance bug found in Apache}
\label{challengingEx}
\end{lstfloat}

\head{Analysis of the program in Listing \ref{challengingEx}} 
The program aims to determine whether a given (target) string is contained within another (source) string. 
If the target string is found in the source string, the program sets the variable \verb+found+ to the index of the target string's first character.
But there is a significant performance flaw in the program: when the target string is at the start of the source string, the run is fast, and the program stops almost instantaneously. 
On the other hand, the run is slower and takes longer to finish when the target string is closer to the end of the source string. 
This is mostly because there will be a significant increase in the number of redundant computations.
The fault is that the initialization statement of the control variable \verb+i+ of the for loop at line 6
should be placed outside the scope of the main while loop just after the initialization of the variable \verb+found+.
The longest run that we reported occurs when the source string has a length of $10^7$ characters, and the target is a single character that is present at the end of the source string. 
In this instance, the program runs for 30 hours before terminating and producing the correct results.

\subsection{Results and analysis}
To handle the performance bug at Listing \ref{challengingEx}, we select two APR tools:
the search-based repair tool GenProg~\cite{legoues2012:genprog} and the semantic-based repair tool FAngelix~\cite{yi2022:speeding}. 
These are general-purpose repair tools for C code that can be used to fix a range of program bugs, including loop program bugs. 
While GenProg successfully generated a plausible patch, FAngelix was unable to produce a plausible one.
To avoid doing repetitive calculations in the original program, GenProg moved the initialization statement of the variable \verb+i+ outside of the for loop at line 6. 
In other words, the program starts with the initialization statement of the variable \verb+i+ in the patched version. 
In this case, the generated patch passes the test cases since  \verb+i+ is no longer being set to 0 every time the loop receives a new character.

To check the validity of the  plausible patch generated by GenProg, we run the tool Daikon
and compare the functional and efficiency predicates obtained for both the original program and the plausible patch. 
Daikon generates the same set of invariants w.r.t. functional variables (i.e., both the original and the patched versions have the same invariants w.r.t. program variables.)
This demonstrates that the patch maintains the functional behavior of the original program. 

Listing \ref{challengingEx} contains four loops: the while loop at line 2, for loop at line 6, while loop at line 9, and for loop at line 15.
To evaluate the efficiency of the original and patched programs, 
it is sufficient to calculate the upper bound on the number of iterations, as the patch does not modify the logic of any of the loops by adding or removing an operation.
That is, each iteration of the four loops in both programs involves the same number of operations.
We therefore add four iteration counters ($cnt_2, cnt_6, cnt_9, cnt_{15})$ to model the efficiency of each loop,
where the index of the counter corresponds to the line number of the loop being analyzed.
For instance, the counter $cnt_2$ is initially set to zero and advanced by one whenever the loop at line 2 is run.
We  make the following observations when analyzing the efficiency predicates for both the buggy and patched versions:

\begin{itemize}
    
    \item Invariants generated for the counter variables $cnt_2$ and $cnt_{15}$ in the buggy and patched versions are the same.   
    This indicates that the patch does not affect the number of times the loops at lines 2 and 15 are iterated.
    \item  The counter variable  $cnt_9$ only advances in the buggy version and results in the invariant  $ cnt_9 \leq 500499$. 
           The fact that the patched version no longer employs the while loop at line 9 is a sign of a major improvement. 

\item Daikon generated the invariant $ cnt_6 \leq 1001$ in the buggy version and invariant $ cnt_6 \leq 501$ in the patched version.
This shows that the loop at line 6 is iterated $50\%$ less times in the patched version than it is in the original code.

\end{itemize}

The aforementioned findings, along with the fact that the derived functional predicates of both the original and patched versions are identical, 
boost our confidence about the validity of the generated patch by the tool GenProg. 

\section{Related Work}

\head{Patch overfitting in APR}
Several solutions have been developed to alleviate the overfitting problem in APR, 
such as symbolic specification inference~\cite{Mechtaev16}, 
machine learning-based prioritization of patches~\cite{Bader}, 
fuzzing-based test-suite augmentation~\cite{Gao}, 
and concolic path exploration~\cite{Shariffdeen21}. 
These solutions rely on limited incomplete test cases and do not guarantee the general correctness of the patches.
Compared to those approaches
that generate test inputs, invariant-based APR automatically 
generates and refines desired invariants that need to be maintained and violated invariants that need to be repaired when modifying code, which makes the approach more reliable than existing repair approaches.

 Modern general-purpose APR tools still rely on symbolic execution or concolic execution \cite{mechtaev2016:angelix,Shariffdeen21}
to discover counterexamples and generate repairs.
However,  these repair approaches manually inspect to determine whether the generated patches are correct or 
identical to developer patches, which could be error-prone. 
Invariant-based APR makes it possible to apply automated 
verification techniques to alleviate overfitting problem and formally and systematically check the accuracy of generated patches 
by comparing them to the developers patches.

\head{Handling performance bugs}
Several attempts have been made to 
detect and repair performance bugs in programs using dynamic, static, and hybrid analysis approaches \cite{SongL17,JinSSSL12,Adrian2013}. 
\cite{Adrian2013} carried out an empirical investigation into performance bugs and presented several efficiency rules for identifying them.
Using dynamic-static analysis techniques, several fix strategies have been developed in \cite{SongL17} to identify and fix performance problems.
However, our method is different from previous studies in that it is a more general and rigorous technique that makes use of program invariant to address loop program performance issues and yield reliable patches. Thanks to program invariants, the original program's efficiency can be systematically compared to the patched version.

\section{Conclusion and Future Work}

We described a novel general-purpose APR system based on the concept of program invariants.
Invariant-based APR holds the promise to handle a wider range of bugs and produce more reliable patches than other APR approaches. 
This is because invariant-based repair systems depend on stronger correctness criteria rather than test suites. 
 We demonstrate the usefulness of leveraging invariants in APR
by developing an invariant repair system for performance defects. 
The preliminary results showed that invariant-based APR can assist in generating valid patches that ensure efficiency improvement without compromising functionality.

\head{Future work}
To complete the line of research initiated here regarding invariant-based APR, we identify the following key directions for future work.
\begin{itemize}
    \item   First and foremost,  we aim to conduct  a thorough empirical analysis to determine how well invariant-based APR handles functional and non-functional defects in programs. 
    This also entails assessing the invariant inference and invariant verification tools that are currently accessible.
    
\item Accurate invariant generation is required to ensure the validity of patches produced by invariant-based APR. 
We conjecture that reachability analyses can aid with this complex computational task and we aim to combine invariant-based APR with program verification tools 
that support both invariant generation and refinement such as CPAChecker and PathFinder.

\end{itemize}

\section*{Acknowledgements}
The author would like to thank Leon Moonen for his insightful discussion  on the idea of the paper.
    This work is supported by the Research Council of Norway through the secureIT project (IKTPLUSS \#288787).

\bibliographystyle{abbrv}
\bibliography{references}

\end{document}